\documentclass[lettersize,journal]{IEEEtran}
\usepackage{amsmath,amsfonts,amssymb}
\usepackage{algorithmic}
\usepackage{algorithm}
\usepackage{array}
\usepackage[caption=false,font=normalsize,labelfont=sf,textfont=sf]{subfig}
\usepackage{textcomp}
\usepackage{stfloats}
\usepackage{url}
\usepackage{verbatim}
\usepackage{graphicx}
\usepackage{xcolor}
\usepackage{cite}
\usepackage{multirow}
\usepackage{hyperref}
\begin{document}

\title{SoundMorpher: Perceptually-Uniform Sound Morphing with Diffusion Model}

\author{Xinlei Niu,~\IEEEmembership{Member,~IEEE,} Jing Zhang,~\IEEEmembership{Member,~IEEE}, and Charles Patrick Martin,~\IEEEmembership{Member,~IEEE}
\thanks{Our demonstration and codes are available at~\url{https://xinleiniu.github.io/SoundMorpher-demo/}}
}

\markboth{Preprint}%
{Shell \MakeLowercase{\textit{et al.}}: A Sample Article Using IEEEtran.cls for IEEE Journals}



\markboth{Preprint version}%
{Shell \MakeLowercase{\textit{et al.}}: A Sample Article Using IEEEtran.cls for IEEE Journals}
\IEEEpubid{0000--0000/00\$00.00~\copyright~2021 IEEE}

\maketitle

\begin{abstract}
We present SoundMorpher, an open-world sound morphing method designed to generate perceptually uniform morphing trajectories. Traditional sound morphing techniques typically assume a linear relationship between the morphing factor and sound perception, achieving smooth transitions by linearly interpolating the semantic features of source and target sounds while gradually adjusting the morphing factor. However, these methods oversimplify the complexities of sound perception, resulting in limitations in morphing quality.
In contrast, SoundMorpher explores an explicit relationship between the morphing factor and the perception of morphed sounds, leveraging log Mel-spectrogram features. This approach further refines the morphing sequence by ensuring a constant target perceptual difference for each transition and determining the corresponding morphing factors using binary search.
To address the lack of a formal quantitative evaluation framework for sound morphing, we propose a set of metrics based on three established objective criteria. These metrics enable comprehensive assessment of morphed results and facilitate direct comparisons between methods, fostering advancements in sound morphing research.
Extensive experiments demonstrate the effectiveness and versatility of SoundMorpher in real-world scenarios, showcasing its potential in applications such as creative music composition, film post-production, and interactive audio technologies.

\end{abstract}


\section{Introduction}

\IEEEPARstart{S}{ound} morphing is a technique to create a seamless transformation between multiple sound recordings. The goal is to produce intermediate sounds that might exist between two sounds and contrasts with simpler mixing or cross-fading between sounds.
Sound morphing has a wide range of applications, including music compositions, synthesizers, psychoacoustic experiments to study timbre spaces~\cite{caetano2011sound,hyrkas2021network}, film post-production, AR/VR, and adaptive audio content in video games~\cite{qamar2020morphino,siddiq2015real}.
Traditionally, sound morphing relied on interpolating the parameters of a sinusoidal synthesis model~\cite{tellman1995timbre,osaka1995timbre,williams2014timbre}. 
High-level audio features in the time-frequency domain can also be controlled to achieve more effective and continuous morphing~\cite{williams2014timbre,brookes2010perceptually,caetano2010automatic,caetano2011sound,roma2020audio,caetano2019morphing}. 
These techniques have quality limitations that rule out applications in inharmonic and noisy sounds such as environmental sounds~\cite{gupta2023towards,kamath2024morphfader}.
Increasing interest in sound generation using machine learning (ML) has delivered techniques for ML-based sound morphing~\cite{nonparallel2021,gupta2023towards,neuralmusic,kamath2024morphfader} that out-perform traditional methods in scenarios such as timbre morphing, audio texture morphing and environmental sound morphing; however, 
we observed critical limitations of these existing sound morphing methods.
Firstly, they are primarily designed for specific morphing methods or tailored to particular application scenarios on task-specific datasets, which limits their ability to broader applications and different scenarios.
Secondly, a lack of sufficient and formal quantitative evaluation limits further analysis of their effectiveness~\cite{gupta2023towards} and impedes fair comparisons with other sound morphing techniques.
Most importantly, these methods typically assume a linear relationship between morphing factors and sound perception, as proposed by~\cite{caetano2012formal}. They achieve smooth morphing by interpolating semantic features linearly between the source and target sounds and gradually varying the morphing factor.
This assumption oversimplifies the complex nature of sound perception, as gradually changing morph factors does not inherently result in smooth perceptual transitions.
\IEEEpubidadjcol
Our goal is to address these limitations by introducing SoundMorpher, \textit{a flexible sound morphing method}, which can be broadly applied to different sound morphing tasks to achieve perceptually coherent morph, ensuring seamless and natural sound transition. In this work we:
\begin{itemize}
    \item Introduce the first open-world sound morphing method based on a pre-trained diffusion model, which integrates typical morph tasks such as static, dynamic and cyclostationary morphing. 
    Unlike~\cite{neuralmusic,gupta2023towards}, SoundMorpher can be broadly applied to various real-world sound morphing tasks without requiring extensive retraining on an additional dataset, paving the way for future sound morphing studies to focus on leveraging pre-trained audio generation models for greater efficiency and versatility.
    \item Propose sound perceptual distance proportion (SPDP), which explicitly connects morph factors and perception of morphed results. This allows SoundMorpher to produce morphing paths with a uniform change in perceptual stimuli, achieving more seamless perceptual transitions compared to existing methods~\cite{kamath2024morphfader}.
    \item Adapt established criteria~\cite{caetano2012formal} for quantitative evaluation, addressing the lack of comprehensive objective quantitative assessment for sound morphing systems~\cite{caetano2019morphing,nonparallel2021,caetano2013musical}.
    \item Provide extensive experiments to demonstrate that SoundMorpher can be effectively applied to various applications in real-world scenarios including timbre morphing, music morphing and environmental sound morphing.
\end{itemize}

\section{Related Work}

In this section, we present a detailed review of related works on sound morphing task and briefly introduce contrasts with other similar tasks.

\textbf{Sound morphing.}
Traditional sound morphing methods are based on interpolating parameters of a sinusoidal sound synthesis model~\cite{tellman1995timbre,osaka1995timbre,williams2014timbre,primavera2012audio}. 
To achieve more effective and continuous morphing, \cite{williams2014timbre,brookes2010perceptually,caetano2010automatic,caetano2011sound,roma2020audio,caetano2011morphing} exploring perceptual spectral domain audio features by digital signal processing techniques, such as MFCCs, spectral envelope, etc.
A hybrid approach has been explored that extracts audio descriptors to morph accordingly and interpolates between the spectrotemporal fine structures of two endpoints according to morph factors~\cite{kazazis2016sound}. 
Machine learning sound morphing methods offer advantages such as high morphing quality by interpolation of semantic representations  within a model instead of traditional audio features.
\cite{nonparallel2021} proposes a non-parallel many-to-one static timbre morphing framework that integrates and fine-tunes the machine learning technique (i.e., DDSP-autoencoder~\cite{engel2020ddsp}) with spectral feature interpolation~\cite{caetano2013musical}.
\cite{neuralmusic} synthesizes music corresponding to a note sequence and timbre, using nonlinear instrument embeddings as timbre control parameters under a pretrained WaveNet~\cite{engel2017neural} to achieve timbre morphing between instruments. \cite{luo2019learning} learns latent distributions of VAEs to disentangle representations of the pitch and timbre of musical instrument sounds.   \cite{tan2020generative} uses a GM-VAE to achieve style morphing to generate realistic piano performances in the audio domain following temporal style conditions for piano performances, which morphs the conditions such as onset roll and MIDI note into input audio. MorphGAN~\cite{gupta2023towards} targets audio texture morphing by interpolating within conditional parameters training the model on a water-wind texture dataset. 
A recent concurrent work~\cite{kamath2024morphfader} uses a pre-trained AudioLDM~\cite{liu2023audioldm} to morph sound between two text prompts. In contrast, we focus on classical sound morphing, where the morphing process is performed directly between two given sounds rather than between text prompts. A key advantage of our method is its ability to provide precise guidance during the morphing process, since the target audio delivers exact information on how the source sound should evolve---something that text prompts cannot always achieve, for example, morphing between two music compositions.

\textbf{Synthesizer preset interpolation.}
A form of sound morphing can be achieved by developing models that compute interpolations in the parameters for a black-box synthesizer~\cite{le2023interpolation,dutoit2023synthesizer,le2024latent}. Unlike classical sound morphing, which perceptually blends two audio files into an intermediate sound, synthesizer preset interpolation treats the synthesizer as a non-differentiable black box, with presets composed of both numerical and categorical parameters. By smoothly interpolating between these presets, the task aims to achieve seamless morphing of synthesized sounds.

\textbf{Text-to-audio editing.}
Text-to-audio editing is the process of using text queries to edit audio. 
With the success of diffusion models in image editing tasks, recent works target zero-shot audio editing with text instructions~\cite{manor2024zero, zhang2024musicmagus,lan2024high} involving tasks such as inpainting, outpainting, timbre transfer, music genre transfer, or vocals removal.

\textbf{Timbre transfer.} 
Timbre transfer is a specific task that aims at converting the sound of a musical piece by one instrument (source) into the same piece played by another instrument (target). This concerns the task of converting a musical piece from one timbre to another while preserving the other music-related characteristics~\cite{comanducci2024timbre,jain2020att,li2024music}.

\textbf{Voice conversion and morphing.}
Voice conversion (VC) involves modifying vocal characteristics of a source speech to match a target speaker, either by using target speeches or text~\cite{li2023freevc,yao2024promptvc,niu24_interspeech,sheng2024voice}. The primary objective of VC is to alter the vocal identity to closely resemble the target voice style while preserving the linguistic content of the source speech. Voice morphing is a broader scope, focusing on blending or transforming one voice into another. This often involves creating an intermediate voice that incorporates characteristics of both source and target voices, allowing for gradual transitions between them~\cite{sheng2024voice}. 

\section{Preliminaries}

\subsection{Sound Morphing}
\label{sec:sound_morphing}

Sound morphing aims to produce intermediate sounds as different combinations of model source sound $\hat{S}_1$ and target sound $\hat{S}_2$~\cite{caetano2010automatic,caetano2011sound}, 
which can be formulated as 
\begin{equation}
    M(\alpha,t) = (1-\alpha(t))\hat{S}_1 + \alpha(t)\hat{S}_2 \label{eqn:morphing}
\end{equation}
Each step is characterized by one value of a single parameter $\alpha$, the so-called morph factor, which ranges between $0$ and $1$. $\alpha = 0$ and $\alpha = 1$ produce resynthesized source and target sounds, respectively.
Due to the intrinsic temporal nature of sounds, sound morphing usually involves three main types: \textit{dynamic morphing}, where $\alpha$ gradually transfers from $0$ to $1$ over time~\cite{kazazis2016sound}, 
\textit{static morphing}, where a single morph factor $\alpha$ leads to an intermediate sound between source and target~\cite{sethares2015kernel}, and \textit{cyclostationary morphing} where several hybrid sounds are produced in different intermediate points~\cite{slaney1996automatic}.

To solve the limitation on previous works that target on expensive perceptual evaluation only, \cite{caetano2012formal} provided three objective criterions to formalize the evaluation for sound morphing methods: 
(1) \textit{Correspondence.} The morph is achieved by a description whose elements are intermediate between source and target sounds, highlighting semantic level transition; 
(2) \textit{Intermediateness.} The morphed objects should be perceived as intermediate between source and target sounds, evaluating perceptual level correlation;
(3) \textit{Smoothness.} The morphed sounds should change gradually from source to target sounds, by the same amount of perception increment. Specifically, the linear assumption proposed by~\cite{caetano2012formal} is that adding the same amount of morph factor should increase the same amount of perception difference. 

In this study, we evaluate SoundMorpher according to the three criterions proposed by~\cite{caetano2012formal} with a series of comprehensive objective quantitative metrics.

\subsection{Latent Diffusion Model on Audio Generation} \label{sec:background_ldm2}

SoundMorpher uses a pretrained text-to-audio (TTA) latent diffusion model (LDM)~\cite{rombach2022high} to achieve sound morphing. This approach offers the advantage of performing various types of sound morphing without the need to train the entire model or use additional datasets. Specifically, we use AudioLDM2~\cite{liu2024audioldm}, a multi-modality conditions to audio model. It employs a pre-trained variational autoencoder (VAE)~\cite{kingma2013auto} to compress audio $x$ into a low-dimension latent space as VAE representations $z$. AudioLDM2 generates latent variables $z_0$ from a Gaussian noise $z_T$ given the condition $C$ and further reconstructs audio $\hat{x}$ from $z_0$ by VAE decoder and a vocoder~\cite{kong2020hifi}.
AudioLDM2 uses an intermediate feature $Y$ as an abstraction of audio data $x$ to bridge the gap between conditions $C$ and audio $x$, named language of audio (LOA). The LOA feature is obtained by a AudioMAE~\cite{huang2022masked,tan2024regeneration} and a series of post-processing formulated as $Y= \mathcal{A}(x)$. 
The generation function $\mathcal{G}(\cdot)$ is achieved by a LDM.
In the inference phase, AudioLDM2 approximates LOA feature by the given condition as $\hat{Y} = \mathcal{M}(C)$ using a fine-tuned GPT-2 model~\cite{radford2019language}. Then generates audios conditioned on the estimated LOA feature $\hat{Y}$ and an extra text embedding $E_{T5}$ from a FLAN-T5~\cite{chung2024scaling} with a LDM as $\hat{x} = \mathcal{G}(\hat{Y},E_{T5})$. We denote the conditional embeddings in AudioLDM2 as $E = \{\hat{Y}, E_{T5}\}$, therefore, the generative process becomes $\hat{x} = \mathcal{G}(E)$.

\textbf{Diffusion Models.}
The LDM performs a forward diffusion process during training, which is defined as a Markov chain that gradually adds noise to the VAE representation $z_0$ over $T$ steps as $z_t = \sqrt{1-\beta_t}z_{t-1} +\sqrt{\beta_t}\epsilon_t$.
where $\epsilon_t \sim N(0,I)$ and noise schedule hyperparameter $\beta_t \in [0,1]$. Therefore, we can derive the distribution of $z_t$ given $z_0$ as $q(z_t|z_0) =\sqrt{\gamma_t}z_0 + \sqrt{1-\gamma_t} \epsilon_t$, where $\gamma_t = \prod_{t=1}^t 1-\beta_t$.
The LDM learns a backward transition $\epsilon_\theta(z_t,t)$ from the prior distribution $N(0,I)$ to the data distribution $z$, that predicts the added noise $\epsilon_t$~\cite{ho2020denoising}.
Following the  objective function of denoising diffusion probabilistic models (DDPM)~\cite{ho2020denoising}, the objective function for training AudioLDM2 is 
\begin{equation}
    \min\limits_{\theta} \mathcal{L}_{\text{DPM}} = \min_\theta [\mathbb{E}_{z_0,E,t\sim\{1,..,T\}}||\epsilon_\theta(z_t,E,t)-\epsilon_t||_2^2] \label{eqn:DPM_loss}
\end{equation}
To reduce computational demands on inference, AudioLDM2 uses 
denoising diffusion explicit models (DDIM)~\cite{song2020denoising}, which provides an alternative solution and enables significantly reduced sampling steps with high generation quality. 
We can revise a deterministic mapping between $z_0$ and its latent STATE $z_T$ once the model is trained~\cite{dhariwal2021diffusion,yang2023impus} by the equation
\begin{equation}~\label{eqn:inv}
    \frac{z_{t+1}}{\sqrt{\gamma_{t+1}}} -\frac{z_t}{\sqrt{\gamma_t}} = (\sqrt{\frac{1-\gamma_{t+1}}{\gamma_{t+1}}} - \sqrt{\frac{1-\gamma_t}{\gamma_t}})\epsilon_\theta(z_t,E,t)
\end{equation}

\begin{figure*}[ht]
    \vskip -0.1in
    \centering
        \centering
        \includegraphics[width=0.78\textwidth]{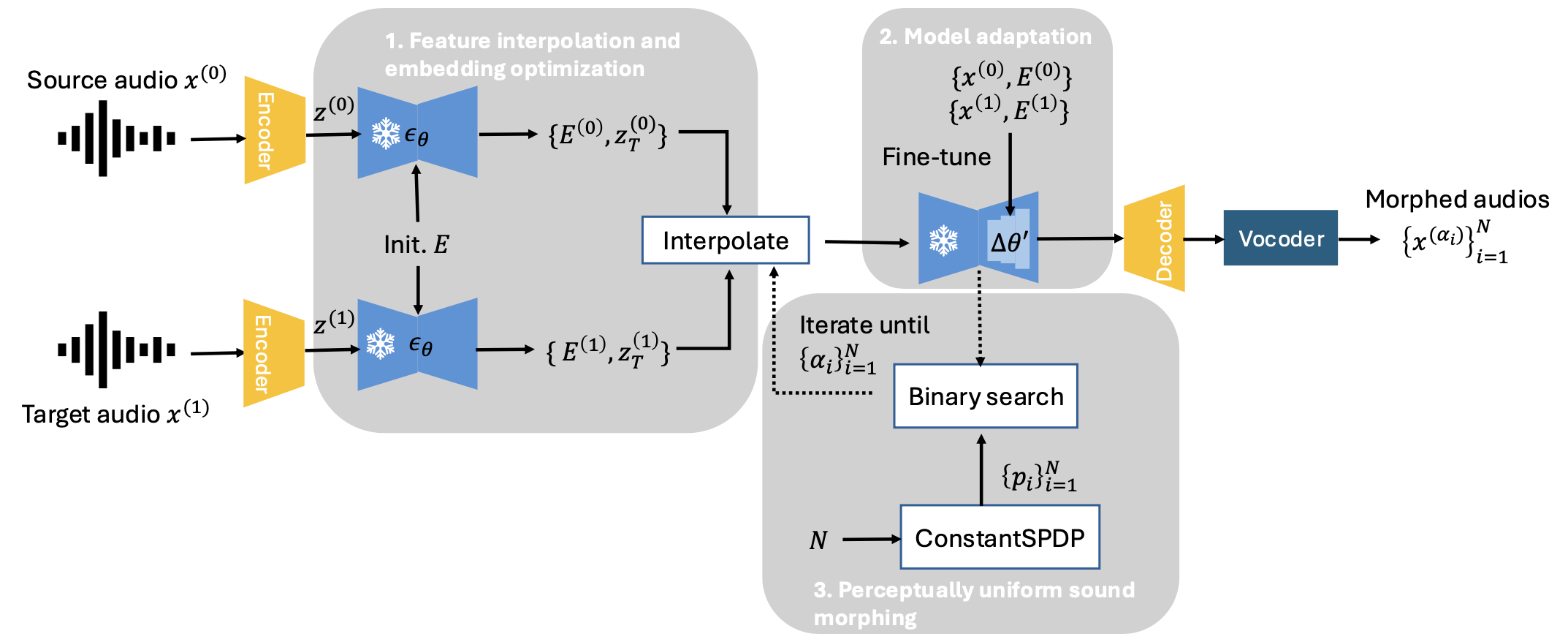}
    \caption{Overview of SoundMorpher pipeline, where the snowflake represents the model parameters are frozen.}
    \label{fig:pipeline}
    \vskip -0.15in
\end{figure*}

\section{Method}
Given a source and target audio pair $\{x^{(0)},x^{(1)}\}$,
sound morphing aims to generate intermediate sounds $x^{(\alpha(t))}$ between the audio pair given morph factors $\alpha \in [0,1]$.  To account for the variation of $\alpha$ over time in Equation~\ref{eqn:morphing}, we discretize the function $\alpha(t)$ where $t\in [0,T]$ into $N$ elements, resulting in a morphed sequence of sounds $\{x^{(\alpha_i)}\}_{i=1}^N$ based on $\{\alpha_i\}^N_{i=1}$.
According to the smoothness criteria proposed by~\cite{caetano2012formal}, a desired sound morphing results should exhibit a \textit{linear perceptual change} as the morph factor $\alpha$ transitions through the sequence ${\alpha_i}^N_{i=1}$.
Therefore, we define $p_i$ to represent the perception of the morphed audio $x^{(\alpha_i)}$ given morph factor $\alpha_i$. However, the relationship $\mathcal{P(\cdot)}$ between morph factor $\alpha$ and perceptual stimuli $p$ is intractable. Our objective is to determine a discrete sequence of morph factors $\{\alpha_i\}^N_{i=1}$ such that the perceptual stimulus difference $\Delta p$ remains constant for each transition.
We formulate the problem as 
\begin{equation}
    p_{i+1} - p_{i} \equiv \mathcal{P}(x^{(\alpha_{i+1})}) - \mathcal{P}(x^{(\alpha_i)}) = \Delta p ~\label{eqn:perceptual}
\end{equation}
where $i \in [1,...,N-1]$.

This formulation is a refined sound morphing problem where, rather than controlling morph factor $\alpha$, we control the constant perception difference $\Delta p$ to find the optimal trajectory with morph factors $\{\alpha_i\}^N_{i=1}$ that will achieve \textit{perceptually uniform sound morphing}. The overall pipeline of SoundMorpher is presented in Fig.~\ref{fig:pipeline} and Algorithm~\ref{alg:overall}.

\begin{algorithm}[h]
\caption{Overall pipeline of SoundMorpher}\label{alg:overall}
\begin{algorithmic}
\REQUIRE A pre-trained AudioLDM2 pipeline including a pre-trained VAE with a encoder $g_{\theta}$ and a decoder $g_{\phi}$, a pre-trained latent diffusion model $\epsilon_\theta$, a pre-trained  T5 model $f_\phi$, and a pre-trained  GPT-2 model $f_\varphi$. Learning rates $\eta_1,\eta_2$. Source and target audios $x^{(0)}$ and $x^{(1)}$. An initial text prompts $y$. perceptually uniform interpolation number $N$. tolerance error for binary search $\epsilon_{tol}$. Number of training steps for text inversion for conditional embedding optimization $T_{inv}$. Number of training steps for model adaptation $T_{adapt}$. LoRA rank $r$, step number of DDIM $T$.

\ENSURE Start morph factor $\alpha_{start} = 0$, end morph factor $\alpha_{end} = 1$. Start perceptual point $p_{start} = [0,1]$, and end perceptual point $p_{end} = [1,0]$.
\STATE \textbf{Initialize:} $z_0^{(0)} = g_\theta(x_0^{(0)})$, $z_0^{(1)} = g_\theta(x_0^{(1)})$; $E^{0} = [f_\phi(y),f_\varphi(y)]$, $E^{1} = [f_\phi(y),f_\varphi(y)]$; 
\STATE \textcolor{gray}{\# Step 1: Text-guided conditional embedding optimization}
\FOR{$i$ {\bfseries from} $1$ {\bfseries to} $T_{inv}$} 
    \STATE Randomly sample time step $t$ and random noise $\epsilon_t \sim N(0,I)$.
    \STATE Adding noise to data $z_t^{(0)} \leftarrow \sqrt{\gamma_t}z_0^{(0)}+\sqrt{(1-\gamma_t)}\epsilon_t,$ \\ $z_t^{(1)} \leftarrow \sqrt{\gamma_t}z_0^{(1)}+\sqrt{(1-\gamma_t)}\epsilon_t$.
    \STATE $E^{(0)} \leftarrow E^{(0)}  -\eta_1\nabla_{E^{(0)}}\mathcal{L}_{\text{DPM}}(z_0^{(0)},E^{(0)};\theta)$.
    \STATE $E^{(1)} \leftarrow E^{(1)}  -\eta_1\nabla_{E^{(1)}}\mathcal{L}_{\text{DPM}}(z_0^{(1)},E^{(1)};\theta)$.
\ENDFOR
\STATE \textcolor{gray}{\# Step 2: Model adaptation with LoRA.}
\FOR{$i$ {\bfseries from} $1$ {\bfseries to} $T_{adapt}$} 
    \STATE Model adaptation with LoRA according to Equation~\ref{eqn:lora1} and Equation~\ref{eqn:lora2} with $\eta_2$ learning rate.
\ENDFOR
\STATE \textcolor{gray}{\# Step 3: Perceptual-uniform binary search with constant SPDP increment}
\STATE Obtaining initial latent states $z_T^{(0)}$ and $z_T^{(1)}$ by Equation~\ref{eqn:inv}
\STATE $p_{list} \leftarrow$ ConstantSPDP$(N,p_{start},p_{end})$  
\STATE $\alpha_{list} \leftarrow \text{BinarySeach}(\alpha_{start},\alpha_{end},p_{list},\epsilon_{tol})$ 
\STATE \textcolor{gray}{\# Interpolate embeddings and latent state}
\FOR{$\alpha$ in $\alpha_{list}$}  
    \STATE $E^{(\alpha)} \leftarrow (1-\alpha)E^{(0)} + \alpha E^{(1)}$
    \STATE $z_T^{(\alpha)} \leftarrow \frac{sin(1-\alpha)w}{sin w}z_T^{(0)} + \frac{sin \alpha w}{sin w}z_T^{(1)}$
    \FOR{$t$ {\bfseries from} T {\bfseries to} 1}
        \STATE $z_{t-1}^{(\alpha)} \leftarrow \sqrt{\gamma_{t-1}}(\frac{z_t - \sqrt{1-\gamma_t}\hat{\epsilon}_\theta^{(t)}(z_t,E^{(\alpha)})}{\sqrt{\gamma_t}}) + \sqrt{1-\gamma_{t-1}}\hat{\epsilon}_\theta^{(t)}(z_t,E^{(\alpha)})$
    \ENDFOR
\ENDFOR
    \STATE $x^{(\alpha)} \leftarrow \text{vocoder}(g_\phi(z_0^{(\alpha)}))$  
\RETURN $\{x^{(\alpha)}\}_{\alpha\in \alpha_{list}}$

\end{algorithmic}
\label{alg1}
\end{algorithm}

In Section~\ref{sec:interpolation} and Section~\ref{sec:model_adaptation}, we introduce feature interpolation and model adaptation with a pre-trained AudioLDM2. This method allows high-quality intermediate morph results to be obtained by interpolating semantic conditional embeddings of source and target sounds according to morph factor $\alpha$.
In Section~\ref{sec:perceptual_uniform}, we explore an explicit connection $\mathcal{P(\cdot)}$ between perceptual stimuli $p$ and morph factor $\alpha$ to achieve perceptually uniform sound morphing, as in Equation~\ref{eqn:perceptual}. This approach enables us to refine the morph factors using binary search to achieve the desired perceptual differences $\Delta p$ for each transition.
In Section~\ref{sec:controll}, we provide extensions of our method on the different morphing methods discussed in Section~\ref{sec:sound_morphing} to show the advantages of perceptually uniform sound morphing.

\subsection{Feature Interpolation}\label{sec:interpolation}
Similar to other sound morphing methods, SoundMorpher achieves sound morphing by interpolating features between source and target audios. In this section, we introduce a method to bridge the source and target audios based on semantic conditional embeddings in a pre-trained AudioLDM2, leading to a controllable sound morphing method by a morph factor $\alpha$.

\textbf{Interpolating optimized conditional embeddings.}
We first introduce a text-guided conditional embedding optimization strategy under a pre-trained AudioLDM2~\cite{liu2024audioldm}, which retrieves corresponding conditional embeddings $E$ of the given audio data. This optimization strategy enables our sound morphing method to handle diverse types of sounds without requiring the model to be retrained on specific datasets. As mentioned in Section~\ref{sec:background_ldm2}, AudioLDM2 accepts two conditional inputs: LOA feature $Y$ and text embedding $E_{T5}$. We denote $E = \{Y,E_{T5}\}$ as the overall conditional embedding inputs for AudioLDM2. The LOA feature $Y$ is an abstraction of audio data which is semantically structured, and $E_{T5}$ captures sentence-level of representations. 
\begin{equation}
\begin{split}
    E^{(0)} = \min_{E} \mathcal{L}_{\text{DPM}}(z_0^{(0)},E;\theta) 
    \text{ and } \\
    E^{(1)} = \min_{E} \mathcal{L}_{\text{DPM}}(z_0^{(1)},E;\theta)
\end{split}
\end{equation}

The optimized conditional embeddings $E^{(0)}$ and $E^{(1)}$ fully encapsulate the abstract details of audios $x^{(0)}$ and $x^{(1)}$.  Due to the semantically structured nature of the conditional embeddings, the conditional distributions $p_\theta(z|E^{(0)})$ and $p_\theta(z|E^{(1)})$ closely mirror the degree of audio variation between the audio pair. To explore the data distribution that conceptually intermediate between $z^{(0)}$ and $z^{(1)}$, we bridge these two distributions through linear interpolation. Specifically, we define the interpolated conditional distribution as $p_\theta(z|E^{(\alpha)}) := p_\theta(z|(1-\alpha)E^{(0)} + \alpha E^{(1)})$, where $\alpha \in [0, 1]$.

\textbf{Interpolating latent state.}
The conditional embedding represents the conceptual abstract of audio. We also wish to smoothly morph the content of the audio pair. 
Following~\cite{song2020denoising}, we smoothly interpolate between $z_0^{(0)}$ and $z_0^{(1)}$ by spherical linear interpolation (slerp) to their starting noise $z_T^{(0)}$ and $z_T^{(1)}$ and further obtained the interpolated latent state $z_T^{(\alpha)}:= \frac{\text{sin}(1-\alpha)\omega}{\text{sin} \omega}z_T^{(0)}+\frac{\text{sin} \alpha \omega}{\text{sin} \omega}z_T^{(1)}$, where $\omega = \text{arccos}(\frac{z_T^{(0)\intercal} z_T^{(1)}}{||z_T^{(0)}|| ||z_T^{(1)}||})$. The denoised latent variable $z_0^{(\alpha)}$ is obtained by applying a diffusion denoising process on the interpolated starting noise $z_T^{(\alpha)}$ and conditioning on the interpolated conditional embedding $E^{(\alpha)}$. The final morphed audio result $x^{(\alpha)}$ is obtained from $z_0^{(\alpha)}$ by the VAE decoder and a vocoder.

\subsection{Model adaptation} ~\label{sec:model_adaptation}
Model adaptation helps to limit the degree of morphed variation by suppressing high-density regions that not related to the given inputs~\cite{yang2023impus}. To enhance the output quality of source and target audios. We follow~\cite{yang2023impus} and use LoRA~\cite{hu2021lora} to inject a small amount of trainable parameters for efficient model adaptation.  We fine-tune AudioLDM2 with LoRA trainable parameters using $z^{(0)}$ and $z^{(1)}$. 
Compared to vanilla fine-tuning approaches, LoRA has advantages in training efficiency with injecting small amount of trainable parameters. The model adaptation can be defined as
\begin{equation}  \label{eqn:lora1}
    \begin{split}
   \min_{\Delta \theta'} \mathcal{L}_{\text{DPM}} (z_0^{(0)},E^{(0)};\theta+\Delta \theta') & + \\ 
    \min_{\Delta \theta'} \mathcal{L}_{\text{DPM}} (z_0^{(1)},E^{(1)};\theta+\Delta \theta') & \text{ s.t. rank} (\Delta \theta') = r  
    \end{split}
\end{equation}
where $r$ represents LoRA rank. 

To achieve high text alignment on inference, we use another LoRA parameters $\Delta \theta_0$ with rank $r_0$ to perform bias correction
\begin{equation} \label{eqn:lora2}
\begin{split}
    \min_{\Delta \theta_0} \mathcal{L}_{\text{DPM}}(z_0^{(0)},\varnothing;\theta+\Delta \theta_0) & + \\
    \min_{\Delta \theta_0} \mathcal{L}_{\text{DPM}}(z_0^{(1)},\varnothing;\theta+\Delta \theta_0) & \text{ s.t. rank} (\Delta \theta_0) = r_0  
\end{split}
\end{equation}
During inference, with $\theta' = \theta +\Delta \theta'$ and $\theta_0 = \theta + \Delta \theta_0$, the noise prediction becomes
\begin{equation}
    \hat{\epsilon}_\theta(z_t,t,E) := w\epsilon_{\theta'}(z_t,t,E)+(1-w)\epsilon_{\theta_0}(z_t,t,\varnothing).
\end{equation}
Although~\cite{yang2023impus} provide a heuristic suggestion for setting the LoRA rank value in the image morphing task; however, we further investigate the relationship between LoRA rank $r$ and method performance in our emprical experiment. 

\subsection{Perceptually Uniform Sound Morphing}\label{sec:perceptual_uniform}

Although interpolating semantic features between the source and target audios enables a smooth transition by leveraging the inherent structure of the semantic space, our ultimate goal is to explore the complex relationship between sound perception and the morph factor to further enhance the smoothness of the morphing sequence. In this section, we propose a method to further refine the morphed results by fixing the perception difference for each transition as constant as in Equation~\ref{eqn:perceptual}.
\textbf{Sound perceptual distance proportion (SPDP).} The relationship between morph factor $\alpha$ and perceptual stimuli $p$ is intractable. 
Our goal is to establish an objective quantitative metric that links $p_i$ and $x^{(\alpha_i)}$ as in Equation~\ref{eqn:perceptual}. This metric should satisfy two key conditions: 
(1) the output $p$ should increase monotonically as $\alpha$ increases; (2) it should accurately represent sound perception and audio descriptive difference between $x^{(\alpha)}$ and $\{x^{(0)},x^{(1)}\}$, ensuring a smooth transition through intermediate states.

To this end, we propose the \textit{sound perceptual distance proportion} between $x^{(\alpha)}$ and $\{x^{(0)}, x^{(1)}\}$. We define $p_i\in \mathbb{R}^2$ as a 2D vector to represent the perceptual proximity of $x^{(\alpha_i)}$ to both $x^{(0)}$ and $x^{(1)}$. Instead of extracting numerous audio features through traditional signal processing techniques, we use log magnitude Mel-scaled spectrogram to capture \textit{perceptual information} and \textit{descriptive aspects} of audio, aligning well with the three objective criteria outlined in Section~\ref{sec:sound_morphing}. The log Mel-spectrogram~\cite{tzanetakis2002musical} provides a pseudo-3D representation of audio signals, with one axis representing time and the other representing frequency on the Mel scale~\cite{stevens1937scale}, while the values denote the magnitude of each frequency at specific time points. This visual representation enables humans to intuitively interpret audio characteristics and descriptive information through its structured depiction of temporal and spectral features. Another advantage of using Mel-spectrogram lies in the Mel filter banks, which map frequencies to equal pitch distances that correspond to how humans perceive sound~\cite{sturm2013introduction,muller2015fundamentals}. 
A logarithmic scaling on the frequency further addresses the imbalance in human perception, which is inherently logarithmic and tends to disproportionately affect low and high-frequency regions. This approach aligns with the principles of psychoacoustics, which suggest that human auditory perception is better represented on a logarithmic frequency scale~\cite{stevens1956direct,davis1980comparison}. 

Denoting $x_{mel}^{(\alpha_i)}$ as the log magnitude Mel-spectrogram of audio $x^{(\alpha_i)}$, the SPDP point $p_i$ of $x^{(\alpha_i)}$ between $x^{(0)}$ and $x^{(1)}$ given $\alpha_i$ is defined as  
\begin{align}\label{eqn:spdp}
    p_i =  & [\frac{||x_{mel}^{(\alpha_i)}-x_{mel}^{(0)}||_2}{||x_{mel}^{(\alpha_i)}-x_{mel}^{(0)}||_2 +||x_{mel}^{(\alpha_i)}-x_{mel}^{(1)}||_2}, \\ 
    ~ & \frac{||x_{mel}^{(\alpha_i)}-x_{mel}^{(1)}||_2}{||x_{mel}^{(\alpha_i)}-x_{mel}^{(0)}||_2 +||x_{mel}^{(\alpha_i)}-x_{mel}^{(1)}||_2}] \nonumber
\end{align}
Equation~\ref{eqn:spdp} calculates the distance proportion of how the morphed sample $x^{(\alpha_i)}$ closes between source $x^{(0)}$ and target $x^{(1)}$, meanwhile the L2 norm measures the distance between two log Mel spectrograms. Therefore, the summation of the two elements in $p$ are always equals to 1. Furthermore, the first element of $p_i$ is strictly monotonically increasing with $\alpha$ while the second element is strictly monotonically decreasing, providing a clear and interpretable relationship between the parameter $\alpha$ and the perceptual proximity of $x^{(\alpha_i)}$ to $x^{(0)}$ and $x^{(1)}$. In summary, the log Mel-spectrogram effectively captures both perceptual and descriptive audio information, offering an elegant solution to satisfy the criterions of smoothness, correspondence, and intermediateness proposed by~\cite{caetano2012formal}.

\textbf{Binary search with constant SPDP increment.}
To produce a perceptually smooth morphing trajectory with a constant perceptual stimuli increment, we use binary search to search for the corresponding $\{\alpha_i\}^N_{i=1}$ based on a constant $\Delta p$ as in Algorithm~\ref{alg:binary_search}. The target SPDP sequence $\{p_i\}_{i=1}^N$ is obtained by interpolation $p_i = (1-\frac{i-1}{N-1})p^{(0)} + \frac{i-1}{N-1}p^{(1)}$,
where the two endpoints are $p^{(0)} = [0,1]^T$ and $p^{(1)} = [1,0]^T$. 

\begin{algorithm}
\caption{Binary search with constant $\Delta p$ }\label{alg:binary_search}
\begin{algorithmic}

\REQUIRE $\alpha_{start}$: start alpha value; $\alpha_{end}$: end alpha value; $N$: interpolate number; source audio $x^{(0)}$; target audio $x^{(1)}$;
\ENSURE $p_{list} = []$, $p_{start} = [0,1]^T$, $p_{end} = [1,0]^T$; $\alpha_{list} = []$, $\alpha_{start} = 0$, $\alpha_{end} = 1$;

\STATE \textcolor{gray}{\# Step 1: Obtain target SPDP points with constant $\Delta p$.}
\STATE \textbf{Procedure} ConstantSPDP$(N,p_{start},p_{end})$
\FOR{$i$ {\bfseries from} $1$ {\bfseries to} $N-1$} 
    \STATE $t  \leftarrow \frac{i}{N - 1}$
    \STATE $p_i  \leftarrow (1-t)\times p_{start} + t\times p_{end}$
    \STATE $p_{list} \leftarrow p_{list} \cup [p_i]$
\ENDFOR
\STATE \textcolor{gray}{\# Step 2: Perform binary search given target SPDP points.}
\STATE \textbf{Procedure} BinarySearch($\alpha_{start}, \alpha_{end},x^{(0)}, x^{(1)}, p_{list}, \epsilon_{tol}$)
    \STATE $\alpha_{list} \leftarrow [\alpha_{start}]$, $\alpha_{cur} \leftarrow \alpha_{start}$;
    \FOR{$p_i$ {\bfseries from} $p_1$ {\bfseries to} $p_{N-2}$}
        \STATE $p_{target} \leftarrow p_i$
        \STATE $\alpha_{t1} \leftarrow \alpha_{cur}$, $\alpha_{t2} \leftarrow \alpha_{end}$
        \STATE $\alpha_{mid} \leftarrow \frac{\alpha_{t1} + \alpha_{t2}}{2}$
        \STATE $p_{mid} \leftarrow SPDP(x^{\alpha_{mid}}, x^{(0)}, x^{(1)})$ 

        \WHILE{$|p_{mid} - p_i| > \epsilon_{tol}$}  
            \IF{$p_{mid} > p_{target}$} 
                \STATE $\alpha_{t2} \leftarrow \frac{\alpha_{t1} + \alpha_{t2}}{2} $
            \ELSE
                \STATE $\alpha_{t1} \leftarrow \frac{\alpha_{t1} + \alpha_{t2}}{2}$
            \ENDIF
            \STATE $\alpha_{mid} \leftarrow \frac{\alpha_{t1} + \alpha_{t2}}{2}$
            \STATE $p_{mid} \leftarrow SPDP(x^{\alpha_{mid}}, x^{(0)}, x^{(1)})$
        \ENDWHILE
        \STATE $\alpha_{list} \leftarrow \alpha_{list} \cup [\alpha_{mid}]$ 
    \ENDFOR
\RETURN $\alpha_{list}$
\end{algorithmic}
\end{algorithm}

\subsection{Controllable Sound Morphing with Discrete $\alpha$ Series}\label{sec:controll}

Once a perceptually uniform discrete morph factor sequence $\{ \alpha_i\}^N_{i=1}$ is obtained, we can further extend it to different sound morphing methods discussed in Section~\ref{sec:sound_morphing}.

\textbf{Static morphing.} To achieve controllable static morphing, we control the target SPDP point $p$,
which represents how the desired output perceptually intermediate between $x^{(0)}$ and $x^{(1)}$. We find the corresponding $\alpha$ value by the binary search with the target $p$ and further obtain a morphed result ${x^{(\alpha)}}$. 

\textbf{Cyclostationary morphing.} To produce $N$ perceptually uniform hybrid sounds between $x^{(0)}$ and $x^{(1)}$, we first obtain $N$ uniform interpolated SPDP points $\{p_i\}^N_{i=1}$.
Then we find corresponding morph factors $\{\alpha_i\}_{i=1}^N$ and further obtain $N$ morphed results $\{x^{(\alpha_i)}\}_{i=1}^N$.

\textbf{Dynamic morphing.}
Dynamic morphing performs sound morphing over time, one challenge is that if the morphing path fails to ensure perceptual intermediateness and audio description correspondence, the resulting sounds may exhibit perceptual discontinuities or unnatural intermediate stages. SoundMorpher produces perceptually uniform sound morphing sequences, which can avoid the problem when performing dynamic morphing. We obtain $N$ interpolated target SPDP points $\{p_i\}^N_{i=1}$ with $\Delta p$. The corresponding morph factors $\{\alpha_i\}_{i=1}^N$ are determined by binary search with the target SPDP points. Each morphed latent variable $z^{(\alpha_i)}$ contributes a segment of length $\frac{K}{N}$, where $K$ is the length of latent variables, producing a latent variable segment $\Tilde{z}^{(\alpha_i)}$ according to index $i$. The final morphed audio sample can be obtained by
\begin{equation}
    x_{morph}= \text{vocoder}(\text{decoder}(\text{concat}(\Tilde{z}^{(\alpha_1)},...,\Tilde{z}^{(\alpha_N)})))
\end{equation}

\begin{table*}[t] 
\footnotesize
\renewcommand{\arraystretch}{1.1}
\renewcommand{\tabcolsep}{1.5mm}
\vskip -0.1in
\caption{Timbral morphing for musical instruments compared to the baseline on different instruments.}
\begin{center}\label{tab:timbre_space}
\vskip -0.1in
\scalebox{1}{
\begin{tabular}{l|l|llllll}
\hline
   Group & Method & FAD $\downarrow$ &  FID $\downarrow$ &  CDPAM$_{T}$ $\downarrow$ & CDPAM$_{mean\pm std} \downarrow$ &  $\mathcal{L}_2^{timbre}$ $\downarrow$ & CDPAM$_{E}\downarrow$\\
\hline
\multirow{2}{*}{Piano $\leftrightarrow$ Guitar }& SMT  & 24.73 & 102.57 & 1.170 & 0.116 $\pm$ 0.074& 1.263 & 0.122\\

                                                & Ours & 5.21 & 41.11 & 0.404 & 0.044 $\pm$ 0.020 & 0.466 & 0.132\\
\hline
\multirow{2}{*}{Harp $\leftrightarrow$ Kalimaba} & SMT &  13.46 & 88.89 & 1.495 & 0.150 $\pm$ 0.117 & 1.355 & 0.182\\
                                                 & Ours &  4.67 & 37.92 & 0.768 & 0.076 $\pm$  0.089 & 0.462 & 0.159\\        
\hline
\multirow{2}{*}{Taiko $\leftrightarrow$ Hihat} & SMT & 8.51 & 131.57 & 2.339 & 0.234 $\pm$ 0.332 & 1.584  &  0.732 \\     
                                               & Ours & 3.32 & 47.59 & 1.314 & 0.131 $\pm$ 0.058 & 0.359  &  0.102 \\
\hline
\multirow{2}{*}{Piano $\leftrightarrow$ Violin}  & SMT & 21.38 & 90.63 & 1.902 & 0.190 $\pm$ 0.069 & 0.558 & 0.217 \\     
                                                & Ours & 3.42 & 20.14 & 0.782 & 0.078 $\pm$ 0.020 & 0.415 & 0.085 \\    
\hline
\multirow{2}{*}{Piano $\leftrightarrow$ Organ}  & SMT  & 21.36 & 63.26 & 1.291 & 0.129 $\pm$ 0.074 & 1.106  &  0.097 \\     
                                                & Ours & 3.29 & 19.73  & 0.233 & 0.023 $\pm$ 0.010 & 0.423 & 0.097\\   
\hline
\end{tabular}}
\end{center}
\vskip -0.25in
\end{table*}

\section{Experiment}
In this section, we show three applications of SoundMorpher in real-world scenarios: \textit{Timbral morphing for musical instruments}, \textit{Environmental sound morphing}, and \textit{Music morphing}. 
\subsection{Evaluation Metric}
We systematically verify SoundMorpher according to the criterions mentioned in Section~\ref{sec:sound_morphing}, with a series of adapted quantitative objective metrics as below
\begin{itemize}
    \item \textbf{Correspondence.}
    We design a metric that computes absolute error for the mid-point MFCCs proportion, namely MFCCs$_{\mathcal{E}}$, for descriptive correspondence. 
    Let $N$ be an odd integer, We define a series of perceptually uniform morphing results $\{x^{(\alpha_i)}\}_{i=1}^N$ with source and target audio $x^{(0)}$ and $x^{(1)}$, where $i$ in the range of $1$ to $N$. The MFCCs$_{\mathcal{E}}$ is computed by
\begin{align}
    \text{MFCCs}_{\mathcal{E}} = \text{abs}( ~~~~~~~~~~~~~~~~~~~~~~~~~~~~~~~~~ &   \\\nonumber
    \frac{||m^{(\frac{N+1}{2})}-m^{(0)}||_2}{||m^{(\frac{N+1}{2})}-m^{(0)}||_2+||m^{(\frac{N+1}{2})}-m^{(1)}||_2} & -0.5)
\end{align} 
    where $m^{(i)}$ represents the extracted MFCCs feature of the $i^{th}$ morphed results in the series $x^{(\alpha_i)}$, $m^{(0)}$ and $m^{(1)}$ represents MFCCs feature of $x^{(0)}$ and $x^{(1)}$. This metric aims to evaluate spectrogram coherent of the midpoint result $x^{(\frac{N+1}{2})}$ between two end points $x^{(0)}$ and $x^{(1)}$, which MFCCs captures \textit{low-level} of audio descriptive information~\cite{davis1980comparison,logan2000mel}. Ideally, we wish the midpoint morphed result contains half-and-half audio description elements on two end points. The larger $\text{MFCCs}_{error}$ indicates the audio description consistency is far away than the midpoint (i.e., 0.5). We extract MFCCs feature with 13 coefficients to compute MFCCs$_{\mathcal{E}}$.
 
    We use \textit{Fréchet audio distance} (FAD)~\cite{kilgour2018fr} and \textit{Fréchet inception distance} (FID)~\cite{eiter1994computing} between morphed audios and sourced audios to verify semantic distribution similarity between morphed samples and sourced audios. Both metrics capture \textit{high-level} of semantic audio descriptive information by extracting audio features from pretrained audio classification  models. Given two sets of audio $X_{mrp}$ and $X_{src}$, where $X_{mrp}$ contains consecutive morphed samples as $X_{mrp} = \{x^{(\alpha_i)}\}_{i=1}^N$, and $X_{src}$ contains sourced audio samples~\footnote{In environmental sound morphing task, we set all sourced sounds in that class as $X_{src}$.}. We calculate FAD and FID values based on audio features extracted by pretrained audio classification models between $X_{src}$ and $X_{mrp}$~\footnote{The FAD and FID metrics are obtained based on \url{https://github.com/haoheliu/audioldm_eval}}.
    \item \textbf{Intermediateness.} CDPAM~\cite{manocha2021cdpam} is a metric designed to evaluate the perceptual similarity between audio signals, which is commonly used in domians such as speech~\cite{manocha2021cdpam}, music~\cite{jacobellis2024machine} and environmental sound~\cite{hai2024dpm}. 
    We use total CDPAM by CDPAM$_{T} =\sum^{N-1}_{i=1} \text{CDPAM}(x^{(\alpha_i)},x^{(\alpha_{i+1})})$ for morph sequence to reflect direct perceptual intermediateness. A smaller CDPAM$_{T}$ indicates the morph sequence exhibits higher perceptual intermediate similarity between consecutive sounds, suggesting intermediate consistency.
    \item \textbf{Smoothness.}
    We calculate the mean and standard deviation of CDPAM along with the morphing path to validate smoothness, as CDPAM$_{mean\pm std}$ = CDPAM$_{mean} \pm \text{CDPAM}_{std}$, where $\text{CDPAM}_{mean} =\frac{1}{N-1} \sum^{N-1}_{i=1} \text{CDPAM}(x^{(\alpha_i)},x^{(\alpha_{i+1})})$, and $\text{CDPAM}_{std} = \sqrt{\frac{1}{N-1}\sum^{N-1}_{i=1} (\text{CDPAM}(x^{(\alpha_i)},x^{(\alpha_{i+1})}) - \text{CDPAM}_{mean})^2}$. 
    In timbre study, we define \textit{timbral distance} as $\mathcal{L}_{2}^{timbre}$ by $\mathcal{L}_2^{timbre} = \frac{1}{N-1}\sum_{i=1}^{N-1} ||q^{(\alpha_{i+1})}-q^{(\alpha_{i})}||_2$, where $q^{(\alpha_{i})}$ represents the corresponding timber point of $x^{(\alpha_{i})}$ in timbre space~\cite{mcadams1995perceptual}. 
    \item \textbf{Reconstruction of perceptual correspondence.} Lastly, we denote CDPAM$_{E}$ that calculate CDPAM between $\{x^{(0)},x^{(1)}\}$ and $\{\hat{x}^{(0)},\hat{x}^{(1)}\}$, where $\hat{x}$ represents resynthesized end points when $\alpha = 0$ and $\alpha = 1$.
\end{itemize}

\subsection{Timbral Morphing for Musical Instruments}

Sound morphing can allow timbral morphing between the sound of two known musical instruments, creating sounds from  unknown parts of the timbre space~\cite{mcadams2013musical,mcadams2017musical}. Timbral morphing for musical instruments involves transitioning between timbres of two different musical instruments to create a new sound. This new sound could possess characteristics of both original timbres as well as new timbral qualities between them, which usually applied to creative arts.
In this experiment, we perform timbral morphing for isolated musical instruments given two recordings of the same musical composition played by different music instruments.

\textbf{Dataset.} 
To ensure high quality of paired compositions on timbral study, we selected 22 paired musical instrument composition samples from demonstration pages of musical timbre transform projects, MusicMagus~\cite{zhang2024musicmagus} and Timbrer~\cite{Timbrer}. The paired samples have durations varies from 5s to 10s, with 16.0kHz and 44.01kHz, which involve 5 groups of instrument pairs: 2 paired samples of piano-violin; 10 paired samples of piano-guitar; 1 paired sample of taiko-hihat; 1 paired sample of piano-organ, and 8 paired samples of harp-kalimaba.

\begin{figure*}[ht]
    \vskip -0.1in
    \centering
    \subfloat[SMT]{%
        \includegraphics[width=0.3\textwidth]{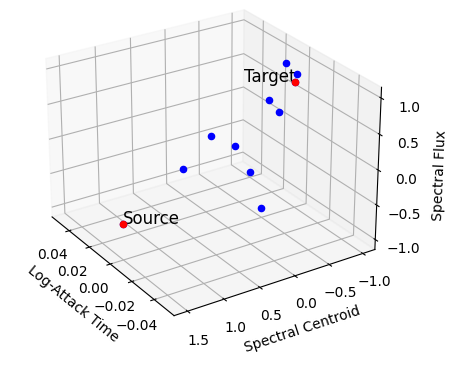}%
        \label{fig:subfig1}%
    }
    \subfloat[SoundMorpher (ours)]{%
        \includegraphics[width=0.293\textwidth]{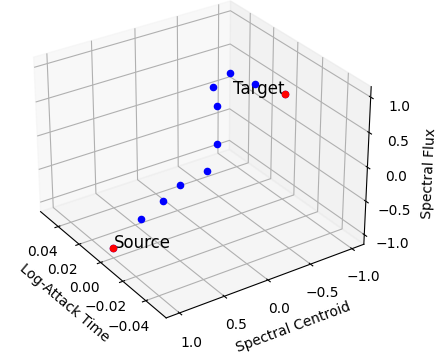}%
        \label{fig:subfig2}%
    }
    \caption{Timbre space visualization of morph trajectories for piano-organ timbre morphing. Compared to SMT, SoundMorpher produces a smoother and continuous morph in the timbre space.}
    \label{fig:timbre}
    \vskip -0.1in
\end{figure*}

\begin{figure*}[ht]
    \centering
        \centering
        \includegraphics[width=\textwidth]{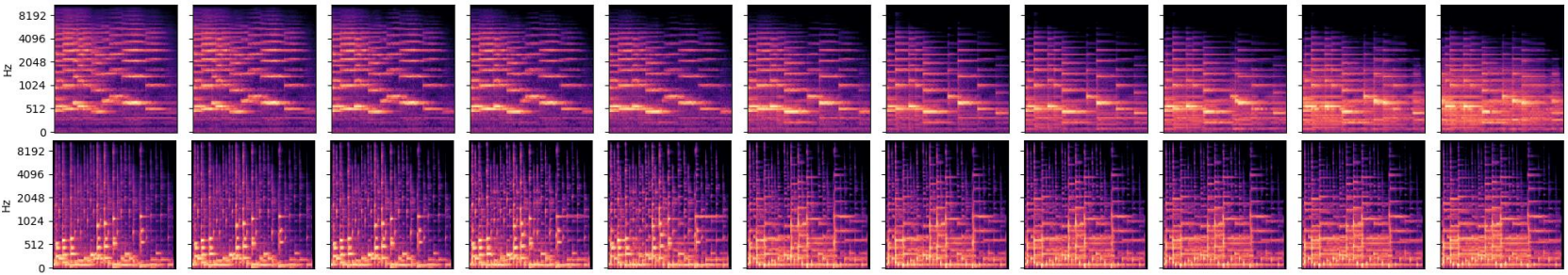}
    \caption{Visualization of timbre morphing for musical instruments with $N = 11$ by SoundMorpher compared to SMT.}
    \label{fig:timbre_vis}
    \vskip -0.15in
\end{figure*}

\textbf{Baseline.} 
We compare with Sound Morphing Toolbox (SMT)~\cite{caetano2019morphing}, a set of Matlab functions targeting on musical instrument morphing. SMT implements a sound morphing algorithm for isolated musical instrument sounds. Since SMT performs sound morphing based on the linear assumption~\cite{caetano2012formal}, we uniformly interpolate 11 morph factors in $[0,1]$.

\textbf{Results and analysis.}
The comparison of our method and the baseline on timbral morphing is in Table~\ref{tab:timbre_space}. Overall, SoundMorpher demonstrates superior morphing quality compared to STM across various metrics, including audio quality, intermediateness, smoothness, and resynthesis quality
~\footnote{Since we perform timbral morphing within the same music composition, MFCCs$_{\mathcal{E}}$ may not a suitable metric under the same musical content. In contrast, we focus on evaluating smoothness and intermediateness.}, when applied to different types of musical instrument timbre morphing. Notably, STM fails in Taiko-Hihat timbral morphing due to significant high reconstruction perceptual error. In contrast, SoundMorpher maintains robustness across different types of musical instruments, making it a more flexible and efficient solution for timbral morphing applications on different types of musical instruments. It is not surprising that SMT has poor performance in this experiment, since the traditional signal processing sound morphing methods are limited to complex or inharmonic sounds~\cite{gupta2023towards}.
Figure.~\ref{fig:timbre} provides a visualization of normalized timbre space, illustrating morphing trajectories generated by SMT and SoundMorpher. The timbre space is defined by three important timbral features: Log-Attack Time, Spectral Centroid, and Spectral Flux~\cite{mcadams1995perceptual,mcadams2013musical}. The SMT trajectory shows distinct steps, indicating that the transitions between each intermediate sound are relatively abrupt. The spacing between the blue points suggests that each step represents a significant change in timbre, which may result in a less smooth perceptual transition between two musical instruments. In contrast, the trajectory produced by SoundMopher demonstrates a smoother curve. The points are more closely spaced, indicating more gradual changes between each intermediate timbre. This suggests that SoundMopher achieves a more continuous and natural-sounding morphing process, with each step being a smaller, more refined adjustment compared to SMT. Fig.~\ref{fig:timbre_vis} provides spectrogram visualization of how SoundMorpher performs timbre morphing.

\subsection{Environmental Sound Morphing}
Environmental sounds are used in video game production to provide a sense of presence within a scene. For example, in video, AR and VR games, sound morphing could enhance user immersion by adapting audio cues to specific visual and interactive contexts. This means that it could be useful to morph between sonic locations, e.g., a city and a park, or between sound effects, e.g., different animal sounds to represent fantasy creatures. In this experiment, we perform cyclostationary morphing with $N=5$ by SoundMorpher across various types of environmental sounds.

\textbf{Dataset.} In this experiment, we firstly use ESC50~\cite{piczak2015dataset} to study the robuteness of SoundMorpher accross different categories of environmental sounds. Then we make a direct comparison with a concurrent method, MorphFader~\cite{kamath2024morphfader}, based on environmental sounds provided in~\footnote{\url{https://pkamath2.github.io/audio-morphing-with-text/webpage/index.html}}.
ESC50~\cite{piczak2015dataset} which consists of 5-second recordings organized into 50 semantic classes which loosely arranged into 5 major categories. We randomly select 4 major categories of scenarios to verify our method, including (1) Dog-Cat (animals voices), (2) Laughing-Crying baby (human sounds), (3) Church bells-Clock alarm (urban noise-interior sound), (4) Door knock-clapping (interior sounds-human sounds). Each category of scenarios contains 25 randomly selected audio pairs, thereby, 100 randomly paired samples in total. To have a fair comparison, we also perform environmental sound morphing based on samples provided by MorphFader, which is sourced from AudioPairBank~\cite{sager2018audiopairbank}.

\textbf{Baseline.} Due to lack of open-sourced direct comparison, we compare SoundMorpher with MorphFader~\cite{kamath2024morphfader} based on the criteria outlined in Section~\ref{sec:sound_morphing}. Although MorphFader didn't release their codes, our evaluation system still can make a fair comparison based on their demonstrated examples.

\begin{table*}[t]
\renewcommand{\arraystretch}{1.1}
\renewcommand{\tabcolsep}{1.5mm}
\caption{Environmental sound morphing with different types of environmental sounds.}~\label{tab:env_sound_morph} 
\begin{center}
\vskip -0.15in
\scalebox{1}{
\begin{tabular}{l|lllllll}
\hline
   Category & FAD$_{\text{category}}$ & MFCCs$_{\mathcal{E}}$$\downarrow$ & FAD$\downarrow$&  FID$\downarrow$  & CDPAM$_{T}$$\downarrow$ & CDPAM$_{mean\pm std}\downarrow$ &   CDPAM$_{E}$$\downarrow$ \\
\hline
Dog $\leftrightarrow$ Cat & 26.08 & 0.081 & 17.77 & 73.92  & 1.293 & 0.323 $\pm$ 0.160 &  0.236  \\
Laughing $\leftrightarrow$ Crying baby & 10.39  & 0.044 & 9.35 & 65.98  & 0.855 & 0.214 $\pm$ 0.077  & 0.289  \\ 
Church bells $\leftrightarrow$ Clock alarm & 68.29  & 0.058 & 22.89 &  75.77  & 2.205  & 0.551 $\pm$ 0.299   &  0.312\\
Door knock $\leftrightarrow$ Clapping & 21.36  & 0.083 & 10.85 & 76.35  &  1.594 & 0.428 $\pm$ 0.220  & 0.321 \\
\hline
\end{tabular}}
\end{center}
\vskip -0.1in
\end{table*}

\begin{figure*}[ht]
    \centering
        \centering
        \includegraphics[width=0.9\textwidth]{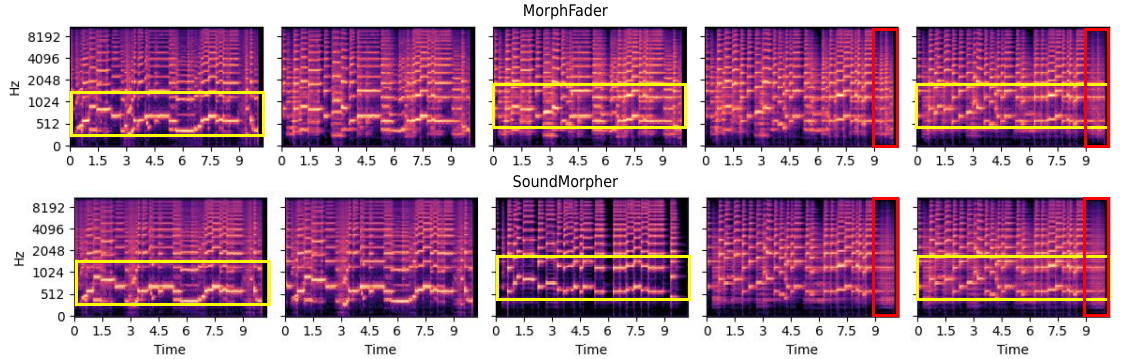}
    \caption{Visualization of spectrogram for morphed results compred with MorphFader. SoundMorpher appears to provide a more seamless and stable morphing process, in which transitions are smoother and the spectral content is more consistent across the morphing stages.}
    \label{fig:model_compare}
    \vskip -0.1in
\end{figure*}

\begin{table}[t]
\caption{Comparison with MorphFader}
\label{tab:morphfader}
\begin{center}
\vskip -0.1in
\begin{tabular}{l|lllll}
\hline
 Method & CDPAM$_{T}$ & CDPAM$_{mean\pm std}$&   MFCCs$_{\mathcal{E}}$\\
\hline
MorphFader   & 0.972 & 0.243 $\pm$ 0.139 & 0.065 \\
SoundMorpher & 0.935 & 0.226 $\pm$ 0.162 & 0.065\\
\hline
\end{tabular}
\end{center}
\vskip -0.3in
\end{table}

\begin{figure*}[h]
    \centering
        \centering
        \includegraphics[width=0.87\textwidth]{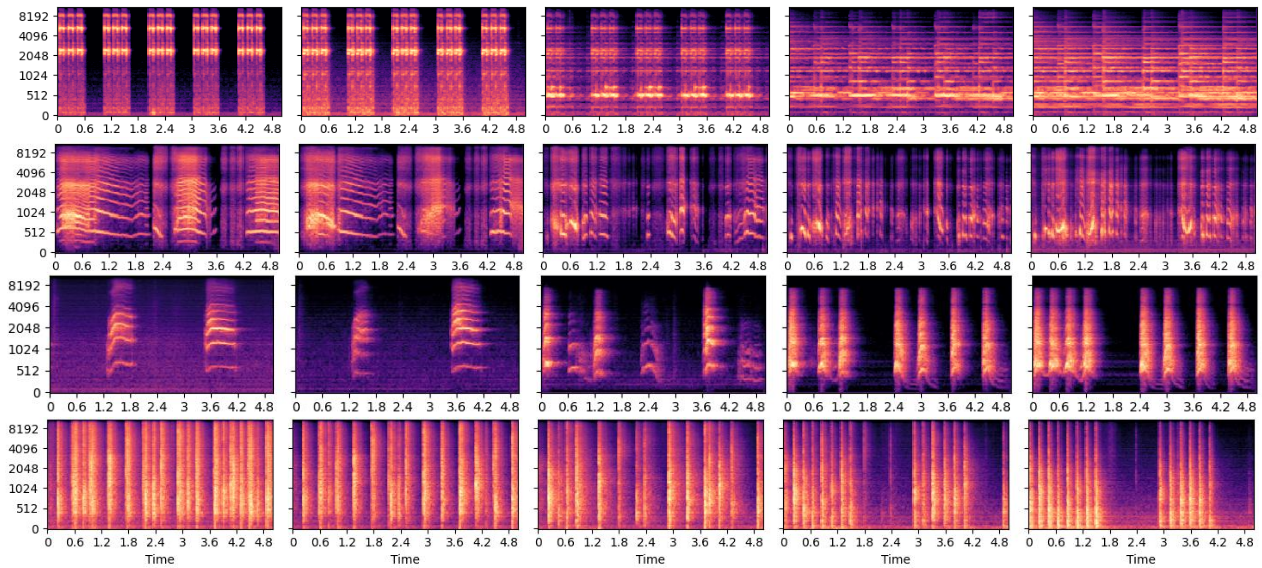}
    \caption{Visualization of environmental sound morphing with $N=5$, from top to bottom: (1) church bells $\leftrightarrow$ clock alarm (2) crying baby $\leftrightarrow$ laughing (3) cat $\leftrightarrow$ dog (4) clapping $\leftrightarrow$ wood door knocking}
    \label{fig:sound_morph_vis1}
    \vskip -0.2in
\end{figure*}

\textbf{Results and analysis.} 
We first apply SoundMorpher on ESC50 dataset to study how SoundMorpher performs on different categories of environmental sounds.
Table~\ref{tab:env_sound_morph} presents the results of applying SoundMorpher to various categories of environmental sounds. To quantify the semantic gap between sound scence classes, we calculate FAD between them as FAD$_{\text{category}}$. The results demonstrate SoundMorpher is capable of effectively morphing a wide range of environmental sounds. However, environmental sounds with a large semantic gap between categories can negatively impact the morphing quality. Additionally, we observe that the quantitative metrics for morphing quality and reconstruction perceptual errors in this experiment are higher than those for the timbre morphing task. One reason is the inherent complexity of environmental sounds, which often involve intricate physical events with significant temporal structure differences and background noises, making them more challenging to morph compared to musical data. Fig.~\ref{fig:sound_morph_vis1} provides spectrogram visualizations on environmental sound morphing. 

We then make a direct comparision with the concurrent method, MorphFader~\cite{kamath2024morphfader}, based on their demonstrated morphed samples in Table~\ref{tab:morphfader}~\footnote{Due to the demonstration page doesn't provide original audio samples, we cannot compute FAD, FID and CDPAM$_{E}$ for this comparison.}. SoundMorpher produces a smoother and intermediate morph than MorphFader with improvement on CDPAM$_{T}$ and CDPAM$_{mean}$. Fig.~\ref{fig:model_compare} provides a direct visual comparison of a morph sequence produced by MorphFader and our method. The yellow rectangles in Fig.~\ref{fig:model_compare} highlight frequency band intensity cross morphing results. The intensity of the frequency bands within the yellow rectangle changes more abruptly for MorphFader, in contrast, SoundMorpher are more stable and consistent across time. This suggests that SoundMorpher maintains better spectral consistency during the morphing process, with smoother transitions between different timbral characteristics.
As red rectangles indicate, MorphFader introduces more abrupt changes at the end of the morphing sequence. There is a noticeable shift in the pattern, indicating less smoothness on transition. In contrast, SoundMorpher shows a more gradual and consistent transition within the red rectangles. The spectral patterns remain more stable and exhibit smoother transitions towards the end of the morphing sequence. Overall, SoundMorpher appears to provide a more seamless and stable morphing process. The transitions are smoother, and the spectral content is more consistent across the morphing stages. 

\subsection{Music Morphing}

Film or game post-production often requires blending or fading between music tracks to seamlessly transition background music in between scenes. Music morphing transitions between two music compositions without cross fading, that is, each moment of the morphed music would be a single composition with elements that are perceptually in between both source and target music, rather than simply blending the two together. Different from timbral morphing, music morphing could ideally be accomplished with compositions from different genres and mixed musical instruments. In this experiment, we use SoundMorpher to perform dynamic morphing on music with $N=15$.

\textbf{Dataset.} Motivated by~\cite{zhang2024musicmagus}, we randomly selected 50 sample pairs from 20 high-quality musical samples available on AudioLDM2~\cite{liu2024audioldm} demonstration page. These 10-second music compositions that span different genres and feature both single or mixed musical instrument arrangements.

\textbf{Baseline.} Due to lack of direct comparison on dynamic music morphing, we set the baseline as SoundMorpher without SPDP binary search, which achieves sound morphing by smoothly interpolate condition embeddings and latent states.  We uniformly sample 15 morph factors in $\alpha\in [0,1]$ as the baseline for comparison. 

\begin{figure*}[ht]
    \vskip -0.2in
    \centering
        \centering
        \includegraphics[width=0.96\textwidth]{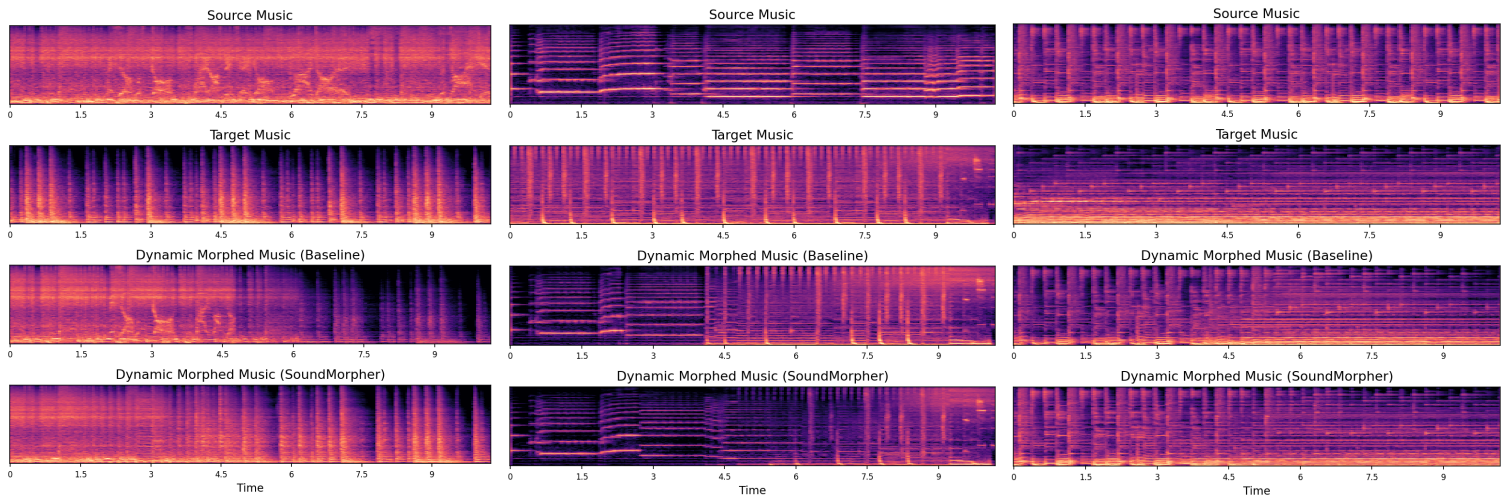}
    \vskip -0.05in
    \caption{Visualization of dynamic morphed music for SoundMorpher and baseline with $N=15$, source music and target music.}
    \label{fig:music_vis1}
    \vskip -0.1in
\end{figure*}

\textbf{Results and analysis.}
Even though this experiment contains morph complex music compositions with different music genres and music instruments, Table~\ref{tab:music_morph} shows our method superiors on perceptual smoothly transiting source music to the target music and ensures correspondence, intermediateness and smoothness compared to the baseline, which further illustrates the advantage of the proposed algorithm of SPDP with binary serach.
Fig.~\ref{fig:music_vis1} visually demonstrates the effectiveness of the proposed dynamic morphing method in transitioning smoothly from source to target music while ensuring perceptual consistency and intermediate transformations. Compared to the baseline results, which exhibit noticeable abrupt changes and spectral discontinuities, SoundMorpher produces gradual and smooth spectral transitions over time. This highlights that simply linearly interpolating semantic features, as done in the baseline method, fails to achieve truly smooth and perceptually consistent sound morphing.

\begin{table*}[t] 
\renewcommand{\arraystretch}{1.1}
\renewcommand{\tabcolsep}{1.5mm}
\caption{Music morphing experimental results, baseline comparison \& ablation study for sound perceptual features} 

\begin{center} \label{tab:music_morph}
\vskip -0.1in
\scalebox{1}{
\begin{tabular}{l|llllllll} 
\hline
Method &  MFCCs$_{\mathcal{E}}$ $\downarrow$ & FAD $\downarrow$ &  FID $\downarrow$  & CDPAM$_{T}$ $\downarrow$ & CDPAM$_{mean\pm std}\downarrow$ & CDPAM$_{E}$ $\downarrow$ \\
\hline
Baseline (Mel-Spec.) & 0.099 & 10.82 & 58.20 & 0.938 & 0.067 $\pm$ 0.065 & 0.158\\
SoundMorpher (Mel-Spec.) & \textbf{0.047} & \textbf{9.85}& \textbf{56.25}  & 0.847 & 0.068 $\pm$ \textbf{0.045} & 0.178  \\
\hline
  Reduced Mel-Spec. (n=2) & 0.187 & 10.31 & 58.57 & 0.793  & 0.056 $\pm$ 0.075  &  0.182  \\ 
  Reduced Mel-Spec. (n=3)  & 0.151 & 10.76 & 59.39 & \textbf{0.779} & \textbf{0.055} $\pm$ 0.079  &  0.151  \\
  MFCCs & 0.053& 10.11 & 57.38  & 0.987  & 0.071 $\pm$ 0.050 &  0.156  \\
 Spectral Contrast & 0.066 & 10.54 & 58.44 & 0.863  &  0.061 $\pm$ 0.071 &  0.155 \\
\hline
\end{tabular}
}
\\ $\star$ where $n$ represents the number of components of PCA for reducing dimension of Mel-spectrogram
\end{center}
\vskip -0.1in
\end{table*}

\subsection{Discussion And Ablation Study}~\label{sec:discussion}

\textbf{Ablation study on sound perceptual features.} We verified the perceptual feature in SPDP in music morphing task. We select alternative music information retrieval features (MIR) including MFCCs with 13 coefficients~\cite{logan2000mel}, and Spectral contrastt~\cite{jiang2002music}. We use principal component analysis (PCA) to reduce the dimensionality of Mel-spectrogram to further capture variation of spectral content over time, which is referred to as reduced Mel-Spec.~\cite{stevens1937scale,casey2008content,jiang2002music}.
Table~\ref{tab:music_morph} shows performance comparisons of SoundMorpher with different features, SoundMorpher with Mel-spectrogram achieves better morphing quality in terms of correspondence and smoothness variation with smaller FAD, FID and CDPAM$_{std}$. While Mel-spectrogram yields higher CDPAM$_{mean}$, CDPAM$_{T}$ and MFCCs$_{\mathcal{E}}$ compared to reduced Mel-Spec. and MFCCs, the differences in metric values are not significant. However, the overall morph quality with Mel-spectrogram is consistently better than other features. This suggests Mel-spectrogram, as a pseudo-3D representation, provide more perceptual and semantic information, which contributes to improve morph quality compared to higher-level features.

\begin{table*}[t] 
\renewcommand{\arraystretch}{1.1}
\renewcommand{\tabcolsep}{1.5mm}

\caption{Experiment and ablation study results on music morphing.} 
\begin{center} \label{tab:ablation}
\vskip -0.1in
\scalebox{1}{
\begin{tabular}{l|ll|lllllll} 
\hline
 Init. text & T = 20  & T =100  &  MFCCs$_{\mathcal{E}}$ $\downarrow$ & FAD $\downarrow$ &  FID $\downarrow$ & CDPAM$_{T}$ $\downarrow$ & CDPAM$_{mean\pm std} \downarrow$  & CDPAM$_{E}$ $\downarrow$ \\
\hline
 Informative & $\checkmark$ &  & 0.047 & 10.21  & 56.62  & 1.213  &  0.086 $\pm$ 0.069 & 0.166  \\
 Informative &  & $\checkmark$  & 0.044 & 10.21 &  56.13 & 1.077 & 0.084 $\pm$ 0.066  & 0.155 \\
  Uninformative & $\checkmark$ & &  0.057 & 10.37  & 55.89  & 1.036  &  0.074 $\pm$ 0.049  & 0.211  \\
 Uninformative & & $\checkmark$  & 0.047 & 9.85 & 56.25  & 0.847  & 0.068 $\pm$ 0.045   & 0.178 \\
\hline
\end{tabular}
}
\end{center}
\vskip -0.2in
\end{table*}

\textbf{Uninformative v.s. informative initial text prompt.} Complex audio usually cannot easily yield precise information to users. For example, it is a challenge for non-professional users to describe the genre of a music. We conduct an ablation study for initial text prompt on music morphing to verify effectiveness of text-guided conditional embedding optimization. We use a general initial text prompt, \textit{`a sound clip of music composition.'}, as an uninformative initial prompt. And we use the given text prompts in AudioLDM2~\footnote{\url{https://audioldm.github.io/audioldm2/}} as informative inital prompts. As in Table~\ref{tab:ablation}, informative initial text prompts may help with resynthesis quality and further improves morph correspondence. Despite the improved resynthesis quality with informative initial text prompts, the results show a decline in morphing intermediateness and smoothness. One possible reason is the better resynthesis quality makes the resynthesis endpoints more distinct (i.e., larger semantic gap), which could lead to slight decline in intermediateness and smoothness. However, the performance difference on initial text prompts is not significant which illustrates effectiveness of conditional embedding optimization.

\textbf{Inference steps.} 
In our experiment, we follow the configuration of~\cite{zhang2024musicmagus} and set DDIM steps to 100. To verify whether DDIM steps affect SoundMorpher performances, we compare with 20 DDIM steps in Table~\ref{tab:ablation}. Larger inference step seems to help for reconstruction quality and slightly imporves morph quality, however, performance differences between inference steps are not significant. This indicates SoundMorpher is robust for inference steps;
therefore, we suggest selecting a suitable DDIM step to trade-off overall binary search algorithm time-consuming and morph quality.

\begin{figure*}[ht]

    \centering
        \centering
        \includegraphics[width=0.86\textwidth]{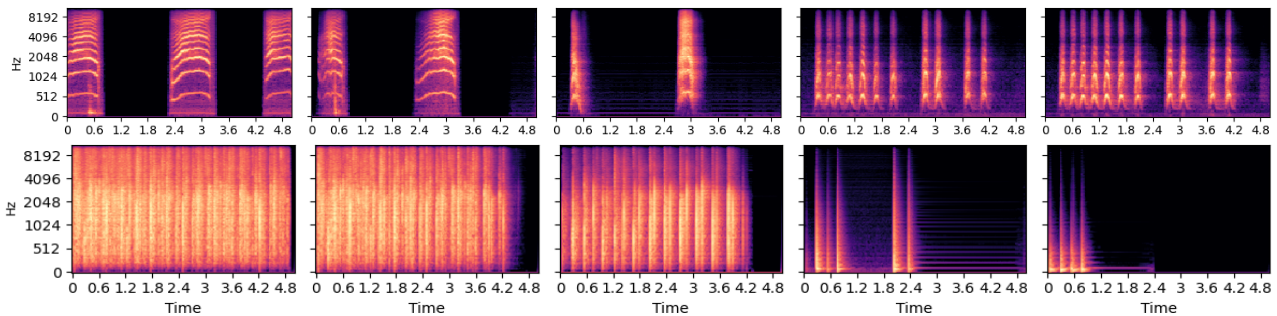}
    \caption{Failure cases for SoundMorpher with $N=5$. The source (left first) and target (right first) sounds that have significant semantic difference in contents, this leads SoundMorpher to  produce abrupt transitions.}
    \label{fig:failure_example}
    \vskip -0.1in
\end{figure*}

\begin{table*}[t]
\caption{Ablation study on model adaptation with different LoRA rank $r$ on music morphing. Rank with $-$ represents results without LoRA model adaptation.}
\label{tab:lora}
\begin{center}
\vskip -0.1in
\scalebox{1}{
\begin{tabular}{l|llllll}
\hline
Rank  & MFCCs$_{\mathcal{E}}$$\downarrow$ &  FAD$\downarrow$ & FD$\downarrow$ &  CDPAM$_{T}$$\downarrow$ & CDPAM$_{mean\pm std}\downarrow$ & CDPAM$_{E}$$\downarrow$\\
\hline
- & 0.073 & 10.38 & 56.02 & 1.052 & 0.085 $\pm$ 0.054 & 0.198\\
4 & 0.047 & 9.85 & 56.25 & 0.847  & 0.068 $\pm$ 0.045  & 0.178 \\ 
8 & 0.059 & 10.01 & 56.35 & 1.035 & 0.073 $\pm$ 0.051 & 0.180 \\
16 & 0.059 & 9.95 & 56.14 & 1.058 & 0.075 $\pm$ 0.052 & 0.169 \\
32 & 0.130 & 10.77 & 59.06 & 0.818 & 0.058 $\pm$ 0.082 & 0.158\\
\hline
\end{tabular}}
\end{center}
\vskip -0.2in
\end{table*}

\textbf{Ablation study on LoRA rank.}
In this experiment, we conduct an ablation study on model adaptation with LoRA on the task of music morphing. We test SoundMorpher with different LoRA ranks as well as SoundMorpher without model adaptation. Following~\cite{yang2023impus}, we train LoRA parameters for 150 steps with 1e-3 learning rate. We also set unconditional bias correction with $r_0 =  2$ for 15 steps with 1e-3 learning rate.
Table~\ref{tab:lora} shows the results of SoundMorpher with different rank size on model adaptation settings. According to Table~\ref{tab:lora}, SoundMorher without model adaptation has obvious performance drop on morphing correspondence compares results with LoRA model adaptation. Even though a higher LoRA rank has a slight improvement on perceptual reconstruction quality, however, SoundMorpher with $r=32$ indicates poor correspondence with large MFCCs$_{\mathcal{E}}$ and large smoothness variance CDPAM$_{std}$. This result indicates that SoundMorpher with higher LoRA rank not lead to a better morphing quality. When $r=4$, SoundMorpher achieves the best performance on smoothness, and correspondence compared to $r=8$, $r=16$ and $r=32$. Therefore, we suggest LoRA rank for model adaptation in SoundMorpher shouldn't be too large.

\textbf{Limitations.} Although SoundMorpher is the first open-world sound morphing method, offering high-quality morphed results and broad applicability, we have identified the following limitations: The current implementation of SoundMorpher based on AudioLDM2 with 16.0kHz sampling rate, which may limit output audio quality. The conditional embeddings optimization only applies to sounds that can be produced by AudioLDM2. Sound examples that close to white noise, such as pure wind blowing used in MorphGAN~\cite{gupta2023towards} are not easily generated by AudioLDM2, which makes the conditional embedding optimization produce low-quality resynthesis sounds. 
Although SoundMorpher produces high-quality sound morphing results, abrupt transitions can occur when the source and target sounds have significant temporal structure differences. A clear example of this is attempting to morph continuous environmental sounds with sounds that contain more silence, as in Fig.~\ref{fig:failure_example}. 
Environmental sounds often consist of discrete and temporally separated events, such as a dog barking or a cat meowing, which have distinct and abrupt characteristics. These are inherently different from the more continuous and harmonically structured nature of music, where elements blend more fluidly over time. As a result, creating smooth transitions between such disjointed environmental sounds can be more challenging, leading to the perception of more abrupt or less natural transitions in the morphing process. 

\section{Conclusion}

We propose SoundMorpher, a open-world sound morphing method base on a pretrained diffusion model that produces perceptually uniform morph trajectories. Unlike existing methods, we refined the sound morphing problem and explored a more complex connection between morph factor and perception of morphed results which offers better flexibility and higher morphing quality, making it adaptable to various morphing methods and real-world scenarios. We validate SoundMorpher with a series of adapted objective quantitative metrics
following three format sound morphing criterions proposed by~\cite{caetano2012formal}. These quantitative metrics may help to formalize future studies for direct comparison on sound morphing evaluation. Furthermore, we demonstrated that SoundMorpher can be applied to wide range of real-world applications in our experiments and conducted in-depth discussions. SoundMorpher also has the potential to achieve voice morphing, as its foundational model AudioLDM2 supports speech generation; however, we leave this exploration for future work.

\bibliography{reference}
\bibliographystyle{IEEEtran}

\end{document}